# Historical Foundation and Practical Guideline for Ferroelectric Switching Kinetic Studies

Yi Liang[1,2], Pat Kezer[2,3], John T. Heron[1,2],*

[1]Department of Materials Science and Engineering,
University of Michigan, Ann Arbor, MI 48109, USA

[2]The Ferroelectronics Laboratory, University of Michigan,
Ann Arbor, MI 48109, USA

[3]Department of Electrical Engineering and Computer Science,
University of Michigan, Ann Arbor, MI 48109, USA

*Corresponding author: jtheron@umich.edu

Electrical measurements of ferroelectric switching kinetics are widely used to probe the dynamics of polarization reversal, yet the influence of the measurement circuit is often underappreciated. In this paper, we show that the interplay between ferroelectric capacitors and circuit elements produces distorted, time-dependent voltage waveforms across the device, particularly in the sub-ns regime. We examine how these circuit contributions affect polarization transients extracted from PUND measurements. The resulting distortions scale with supply voltage, capacitor dimensions, and lumped circuit elements, but are not accounted for in conventional experimental analyses or analytical model fitting. We then critically assess existing nucleation and growth models and show that neglecting the time-varying voltage profile can lead to unphysical interpretations of switching kinetics, most notably in the extracted growth dimensionality represented by the Avrami exponent. Finally, we outline guidelines for future studies, emphasizing the need for direct voltage monitoring and circuit-aware de-embedding, as well as modeling frameworks that incorporate voltage-dependent nucleation and growth rates based on intrinsic material parameters.

## 1. Introduction

In the 1970s, initial development of ferroelectric materials into thin films began due to the accompanied need for integrated non-volatile memories and the maturation of thin film processing techniques[1,2]. Since then, the number of ferroelectric thin films has grown immensely along with their potential to solve major challenges in a variety of application spaces, notably emergent logic-in-memory, neuromorphic, and other AI computing technologies. In such a technology, a high device density (cell size in sub-100 nm), low power dissipation, and fast switching speed (sub-ns) [3] are needed to surpass current CMOS implementations. From these and the realization that computing can be done at the fundamental limit of materials behavior, the need to understand the dynamics of polarization switching in thin films has become imperative, in particular as the volume of the ferroelectric material approaches sub-micron scales. In more recent years, many reports of switching dynamics have been made through electrical measurement of the dielectric displacement current from micron-scale capacitors and in the sub-ns regime[4–10].

The common implementation of these electrical measurements uses positive-up-negative-down (PUND) voltage pulse trains[11] (**Figure 1a**) that are applied across the ferroelectric capacitor while the resulting displacement current from polarization switching is measured directly or across a reference resistor. When two identical pulses (such as the P and U pulses) are applied to the ferroelectric capacitor, it is conventionally believed that the parasitic background is measured by the second pulse and can be subtracted from the total signal measured from the first pulse to obtain the true switching transient. Embedded in the switching transient is the nucleation and growth process of reversed domains, which are interpreted by fitting the transients to the analytical models that describe the transformed fraction as a function of time in terms of material parameters and voltages (**Figure 1b**). The earliest kinetic model of ferroelectric switching, by Ishibashi and Takagi[12], was adapted from the traditional crystal growth theory of Kolmogorov and Avrami[13–15] and is now commonly known as the Kolmogorov-Avrami-Ishibashi (KAI) model. Later, several alternative analytical models, such as nucleation-limited switching (NLS) model[16] and simultaneous non-linear nucleation and growth model (SNNG) model[17], have been proposed and commonly used to better capture nuances of the switching transient profiles. However, all of these theories assume a constant voltage is across the ferroelectric capacitor, i.e. perfect square pulses with 0 s rise/fall time and no circuit loading effects, which enables the extrapolation of voltage dependent parameters using Merz law[18].

For decades, the utilization of PUND measurements plus simple model fitting has become the norm as it not only speculates the fundamental switching mechanisms (the competition between nucleation and growth), but also can be used to feedback synthesis and device operation, as the effect of scaling, waking up cycles and fatigue[17,19,20] are reflected in changes in nucleation density, domain propagation rate, and ultimately the fit parameters extracted by analytical models. While modification of models can provide a fit to experimental data, the underlying physics of any modification and the methodology of data collection need careful justification. For example, it has long been pointed out that the $RC$ time constant of the measurement circuit[8] and significant ferroelectric switching (displacement) current would distort the voltage waveform intended for switching the polarization [21,22] (**Figure 1c**), yet most studies don't even measure the voltage across the capacitor and assume the ideal waveform in PUND measurements. Moreover, the analysis of the transients with current analytical models does not allow the incorporation of a time-dependent voltage. Consequently, it becomes questionable whether these models can still predict ferroelectric response under an actual application circuit. These problems prevent accurate extraction of materials behavior and the later design of microelectronic components where dynamic and mixed signal waveforms are used.

Here we summarize ferroelectric switching transient measurements, the nucleation and growth analysis of ferroelectric switching, discuss the limitations to common methodologies, and inspire solutions to them. This review is organized as follows. In the second section, we revisit the basics of the PUND method and common practices in the extraction of the ferroelectric switching transient. Then we examine and quantify the errors in PUND measurements if circuit effects are overlooked. In the third section, we discuss the account of prevalent nucleation and growth models used in the analysis of switching transients, with an emphasis on the underlying assumptions of each. Overlooked circuit effects can also lead to misuse of kinetic models beyond their range of validity and overinterpretation of experimental data. Finally, we provide a perspective for new kinetic nucleation and growth models that account for ultrafast and time-dependent waveforms (**Figure 1d**).

## 2. Experimental extraction of switching kinetics

### 2.1 Free and bound charges and the electric displacement vector

To measure and interpret electrical measurements of ferroelectric switching transients, we must have a way to connect the displacement current measured to the dielectric response. An ideal ferroelectric material, one without leakage current or space charge, would be described in terms of surface bound charge $\sigma_b = \boldsymbol{P}_{tot} \cdot \hat{\boldsymbol{n}}$ and volume bound charge $\rho_b = -\nabla \cdot \boldsymbol{P}_{tot}$ that result from surface termination or bulk gradients in the polarization ($\boldsymbol{P}_{tot}$) vector field. Here bound charges are charges bound to atoms or molecules in the ferroelectric, and the total polarization includes linear dielectric component from reversible dipoles ($\boldsymbol{P}_{DE}$) and nonlinear switchable ferroelectric component ($\Delta \boldsymbol{P}$). The electrical measurement of ferroelectric switching transients requires that metal electrodes sandwich the ferroelectric forming a parallel plate capacitor. The metal electrodes provide free charge ($\rho_f$) that can compensate or screen bound surface charge. The total charge density enclosed in the system is then $\rho = \rho_b + \rho_f$. A connection between the applied electric field and the total charge density in a ferroelectric is given by Gauss's law: $\epsilon_0 \nabla \cdot \boldsymbol{E} = \rho$. Then by substitution of the above expressions, we have

$$\nabla \cdot (\epsilon_0 \boldsymbol{E} + \boldsymbol{P}_{tot}) = \rho_f \tag{1}$$

where the vector sum in the parentheses defines the electric displacement vector $\boldsymbol{D}$, i.e. $\boldsymbol{D} = \epsilon_0 \boldsymbol{E} + \boldsymbol{P}_{tot} = \epsilon_0 \boldsymbol{E} + \boldsymbol{P}_{DE} + \boldsymbol{P}$, which is key to understanding and modeling capacitance, displacement current density $(d\boldsymbol{D}/dt)$, and therefore ferroelectric circuits. With $\epsilon_0 \boldsymbol{E} + \boldsymbol{P}_{DE} = \epsilon_{DE} \boldsymbol{E}$, $\epsilon_{DE}$ being the linear dielectric constant of the ferroelectric capacitor, the displacement current density that flows through the system is:

$$\boldsymbol{J} = \frac{d\boldsymbol{D}}{dt} = \epsilon_{DE} \frac{d\boldsymbol{E}}{dt} + \frac{d\boldsymbol{P}}{dt} \tag{2}$$

or in a projected scalar form:

$$I = C_{DE} \frac{dV}{dt} + A \frac{dP}{dt} \tag{3}$$

where $C_{DE}$ is the equivalent linear dielectric capacitance of a ferroelectric. According to **Equation 3**, the metal-ferroelectric-metal (MFM) capacitor structure is usually modeled as a linear (first term) and a nonlinear (second term) capacitor in parallel (**Figure 1a**).

**2.2 Ideal PUND measurement**

To properly extract the ferroelectric switching transient, positive-up-negative-down (PUND) pulse train method was proposed in 1989[11]. The ferroelectric is first reset to a uniformly polarized state before P (switching) and U (non-switching) pulses (**Figure 1a**). The P pulse is applied and excites a total P pulse current:

$$I_P = C_{DE} \frac{\partial V_P(t)}{\partial t} + A \frac{\partial P(V_P(t))}{\partial t} \tag{4}$$

The U pulse takes place after the ferroelectric capacitor is fully switched and only the linear capacitor is charging, generating a pure non-switching current:

$$I_U = C_{DE} \frac{\partial V_U(t)}{\partial t} \tag{5}$$

As in an ideal scenario P and U pulses are assumed to be identical, the current purely from ferroelectric polarization switching can be extracted by subtracting **Equation 5** from **Equation 4**. Integrating the ferroelectric current with respect to time gives the change in polarization versus time as driven by the P pulse.

$$\Delta P(t) = \frac{1}{A} \int_0^t (I_P - I_U) \, dt \tag{6}$$

The N and D pulses have reversed polarity to P and U pulses and follow similarly under the given assumptions. The ideal PUND method works for pulse trains and triangular waveforms and is the predominant methodology used today to measure remnant polarization and *P(V)* hysteresis loops. Generally, switching and non-switching pulses are also used to read out how much polarization is left unswitched by the prior waveforms after the reset pulse. This measurement scheme is ubiquitous in testing devices for neuromorphic computing where polarization values are mapped to the programming pulses.

## 2.3 Non-ideality in PUND measurements

### 2.3.1 Irreducible circuit effects

In a real measurement circuit, the actual voltage waveform across the ferroelectric capacitor can deviate from the output waveform of the source for the following reasons. First of all, the rise time of the signal generator is finite. The internal output impedance of the signal generator coupled with the capacitive nature of the ferroelectric material results in an RC-like voltage response across the ferroelectric capacitor that further slows down the voltage rise. Moreover, during ferroelectric switching, significant current is produced and flows through series resistors of the oscilloscope. Consequently, the voltage across the ferroelectric capacitor is always lower than the set voltage during switching[21]. One way to mitigate these effects is to use very low pulse amplitude[23], so that the switching time is much longer than the rise time and RC constant, and the ferroelectric switching current is small enough to circumvent considerable voltage drop across the series circuit resistance. However, this scheme does not align with the need of pushing the limit of ferroelectric switching down to the sub-ns regime, and it is unclear how much voltage distortion can significantly impact polarization transients. In this session, we analyze the circuit effects on polarization transients with actual voltages measured across a metal-ferroelectric-metal (MFM) capacitor structure.

We first assume the MFM structure only includes the intrinsic nonlinear and linear capacitor describing an ideal ferroelectric capacitor. The size of these circuit effects is dependent on the device dimension and voltage amplitude. To obtain the correct $V_P$ and $V_U$ values, an active probe is used to monitor the voltage between the top electrode and the ground ($V_{top}$), and a GSG probe for voltage between bottom electrode and the ground ($V_{bottom}$) as well as the current transient. The actual voltage across the ferroelectric capacitor $V_{FE}$ is calculated by $V_{top} - V_{bottom}$, for both P and U pulses. The equivalent circuit is shown in **Figure 2a**. According to **Equation 5**, The linear capacitance $C_{DE}$ is derived by a linear fit between $\frac{\partial V_U}{\partial t}$ and $I_U$. The error in polarization is then calculated by

$$\frac{\partial P_{err}}{\partial t} = \frac{C_{DE}}{A}\left(\frac{\partial V_U}{\partial t} - \frac{\partial V_P}{\partial t}\right) \tag{7}$$

Integrating with respect to time gives

$$P_{err}(t) = \frac{C_{DE}}{A}\left(V_U(t) - V_P(t)\right) \tag{8}$$

Here we use a 10-μm Hf$_{0.5}$Zr$_{0.5}$O$_2$ (HZO) capacitor[24] as an example (**Figure 2b**). The rising edge of the P pulse is more skewed than the U pulse, because the ferroelectric current is not fully separable from the dielectric current. During ferroelectric switching, the voltage drop compared to a set square pulse can reach about half of the supply voltage. The error in the transient polarization first increases to a peak value (~9% in this case) and eventually falls back to zero. Notably, while the measured switched polarization is not affected, the corrected polarization transient here shifts to smaller time scale, which is usually interpreted as shorter incubation time. Furthermore, as will be discussed in section 3, the dynamic change in the polarization leads to errors in model fitting and interpretation. Interestingly, when different supply voltages are applied, the actual P pulse voltages across the capacitor are clamped to a similar value (~1.4 V), with higher supply voltage resulting in a steeper slope in the voltage profile in general. As a consequence, the errors in the transient polarization scale with supply voltages (**Figure 2c**). Similar analysis is performed on devices of diameters from 1 to 30 μm. While only the voltage waveform across the 1-μm capacitor seems to preserve the square pulse shape, the voltages across larger capacitors are clamped to a similar value (~1.5 V) during switching. For all the observed device sizes, because of non-identical P and U pulses, the errors in transient polarization reach a peak value of ~ 5-9% (**Figure 2d**).

We note that the quantification here is done on HZO with moderate dielectric constant ($\epsilon_r$~25) and driving fields ($\leq$ 2.3 × coercive field). If PUND measurements are performed on materials with much higher dielectric constants (e.g. titanates) or when much higher driving fields (e.g. for wurtzite ferroelectrics) are used, the actual waveform will deviate more from the programmed waveform. Higher errors can occur in the polarization transients. Furthermore, even though the waveform is expected to be less distorted in sub-micron devices, as the switching time also shortens, the influence of the rise time may in turn dominate. Therefore, monitoring the actual voltage waveform becomes necessary when extracting ultrafast switching kinetics.

*2.3.2 Parasitic linear components*

In most cases, the MFM structure between the two measurement nodes (probes) includes parasitic linear components besides the ideal ferroelectric capacitor due to the selected electrode materials and device fabrication process (**Figure 3a**). To define a ferroelectric capacitor cell with uniform distribution of electric field, capacitor island instead of continuous ferroelectric film needs to be fabricated[25]. The isolation materials (e.g. SiO$_2$) separating the top and bottom electrodes surround the ferroelectric island and equivalently form a parasitic capacitor in parallel ($C_p$). With the parallel capacitor, the total linear capacitance is $C'_{DE} = C_{DE} + C_p$. Substituting $C_{DE}$ with $C'_{DE}$ in **Equation 8** maintains the validity of our previous analysis on the polarization error, as the parallel capacitor shares the same voltage value across the ferroelectric capacitor. However, the top and bottom electrodes may introduce extra series resistance in addition to the internal resistances of the signal generator and the oscilloscope. Time-varying current flowing through the electrodes also introduces series inductance. The series inductors ($L_s$) and resistor ($R_s$) induce a voltage drop across the ferroelectric capacitor, which makes the P and U pulses become further inequivalent. The true voltage across the ferroelectric capacitor for P and U pulses now becomes

$$V_P = V_P^{meas} - \frac{\partial I_P}{\partial t}L_s - I_P R_s \tag{9}$$

$$V_U = V_U^{meas} - \frac{\partial I_U}{\partial t}L_s - I_U R_s \tag{10}$$

Applying the new voltages to **Equation 4** and **5**, and using the total dielectric capacitance $C'_{DE}$,

$$I_P = C'_{DE} \frac{\partial(V_P^{meas} - \frac{\partial I_P}{\partial t}L_s - I_P R_s)}{\partial t} + A\frac{\partial P}{\partial t} \tag{11}$$

$$I_U = C'_{DE} \frac{\partial(V_U^{meas} - \frac{\partial I_U}{\partial t}L_s - I_U R_s)}{\partial t} \tag{12}$$

The error in polarization is given by

$$P_{err}(t) = \frac{C'_{DE}}{A}\left(V_U^{meas}(t) - V_P^{meas}(t) + L_s\left(\frac{\partial I_P}{\partial t} - \frac{\partial I_U}{\partial t}\right) + R_s(I_P(t) - I_U(t))\right) \tag{13}$$

The above equations suggest that with the series inductor and resistor, $C'_{DE}$ can no longer be derived by a simple linear fit as described by Equation 5. In order to recover true ferroelectric polarization transients, individual values of the parasitic elements embedded in the MFM structure need to be measured. This can be achieved by using on-chip "OPEN" and "THRU" structures. Moreover, from an application aspect, the decrease in the switching pulse voltage can also compromise the switching speed of the ferroelectric. Therefore, device geometry should be carefully considered to minimize parasitic resistance and inductance, such as reducing ground loops with the use of an on-chip transmission line[6,8,24], reducing the length of single-end leads, as well as reducing the length or increasing the width of resistive electrodes (in the case of oxide electrodes as opposed to metals). The design of the on-chip ferroelectric cell structure and the de-embedding of the parasitic elements were largely overlooked in previous studies of ferroelectric switching kinetics and warrant future attention.

*2.3.3 Errors from material deficiencies*

*Dead layers*

Due to the mismatch in atomic structure, roughness, and optimal growth conditions between ferroelectric materials and electrode, dead layers, which are essentially non-polar phases, can form at the interface as a result of strain and defects[26–29]. The dead layer can be modeled as a linear capacitor ($C_s$) in series with the ferroelectric capacitor (**Figure 3b**). In this case, during P pulse, as the charge ($Q$) on the ferroelectric capacitor equals that on the series capacitor, i.e. $Q = PA + C_{DE}V_P = C_s(V_P^{meas} - V_P)$, the voltage across the ferroelectric capacitor can be obtained as

$$V_P = \frac{(C_s V_P^{meas} - PA)}{C_{DE} + C_s} \tag{14}$$

Substituting **Equation 14** to **Equation 4** gives the P pulse current,

$$I_P = A\frac{C_{DE}}{C_{DE} + C_s}\frac{dP}{dt} + \frac{C_{DE}C_s}{C_{DE} + C_s}\frac{dV_P^{meas}}{dt} \tag{15}$$

During U pulse, the circuit only includes two linear capacitors in series, and the current is calculated as

$$I_U = \frac{C_{DE}C_s}{C_{DE} + C_s}\frac{dV_U^{meas}}{dt} \tag{16}$$

Therefore, the transient polarization should be

$$\Delta P(t) = \frac{C_{DE} + C_s}{AC_s}\int_0^t (I_P - I_U)\,dt + \frac{C_{DE}}{A}(V_U^{meas} - V_P^{meas}) \tag{17}$$

Comparing **Equation 17** to **Equation 6**, the error in the transient polarization is now

$$P_{err}(t) = \frac{1}{A}\frac{C_{DE}}{C_s}\int_0^t (I_P - I_U)\,dt + \frac{C_{DE}}{A}(V_U^{meas}(t) - V_P^{meas}(t)) \tag{18}$$

Note that because the error now includes an integral of the current difference between P and U pulses, the error will not return to zero at the end of the switching, in contrast to all the cases discussed in previous sessions. The actual polarization of the ferroelectric material is higher than what is measured by conventional PUND method if a dead layer is present. Therefore, dead layers can affect both the switching kinetics and the remnant polarization. More generally, the model of a ferroelectric capacitor in series with a linear capacitor is useful not only in simulating dead layers, but also in understanding depolarization field[30] and the recently discovered negative capacitance[29,31].

*Leakage*

Point and extended defects in oxide and nitride ferroelectrics often lead to leakage that degrades ferroelectric memory, energy efficiencies and endurance. In a vertical MFM structure, scaling down the ferroelectric films to nanometers thickness can produce lower resistance and more severe leakage. Leaky ferroelectrics may exhibit semiconducting behavior with nonlinear resistivity. For simplicity we model the leakage with a nonlinear resistor ($Rp$) in parallel with the ferroelectric capacitor whose value can be dependent on the voltage across it (**Figure 3b**). Then the currents during P and U pulses will have additional terms $\frac{V_P^{meas}}{R_p(V_P^{meas})}$ and $\frac{V_U^{meas}}{R_p(V_U^{meas})}$ respectively. The error in transient polarization becomes

$$P_{err}(t) = \frac{C_{DE}}{A}\left(V_U^{meas}(t) - V_P^{meas}(t)\right) + \int_0^t \left(\frac{V_U^{meas}}{R_p(V_U^{meas})} - \frac{V_P^{meas}}{R_p(V_P^{meas})}\right) dt \qquad (19)$$

In this equation, the integral of voltage difference is involved and similar to the series capacitor case, the measured switched polarization appears to be lower if leakage occurs.

The parasitic components introduced into the MFM structure by dead layers and leakage cannot be excluded by standard de-embedding methods but can only be mitigated through synthesis optimization. In addition, they introduce local defects and roughness variations that can act as pinning or nucleation sites, as well as additional Joule heating that modifies the energy landscape of the polarization states. To enable rigorous comparison and interpretation on the dynamic behaviors across different material systems and device architectures, these effects should be clarified or removed.

## 3. Modeling of switching kinetics

Once ferroelectric polarization transient is measured by PUND method, a model of the phase evolution (say a classical nucleation and growth model) can be used to provide fundamental microscopic insight into the mechanism of the evolution with the extraction of material parameters. Besides, as the evolution of ferroelectric polarization is nonlinear, presenting a sigmoidal shape, compact analytical models are needed for large-scale circuit modeling. As a simplified example, substituting time-dependent polarization into **Equation 3** and iteratively solving the differential equation can simulate ferroelectric circuit response (current and real-time voltage) from an initial stimulus (assumed voltage waveform)[24]. In this session, we review the development and the limitation of analytical nucleation and growth models for ferroelectric switching kinetics, discuss associated problems due to the ignorance of distorted waveforms, and provide future directions of improvement in modeling.

### 3.1 Classical nucleation and growth models: The JMAK and KAI models

The KAI (Kolmogorov-Avrami-Ishibashi) model borrowed the theory from the nucleation and growth model for crystallization, the Johnson-Mehl-Avrami-Kolmogorov (JMAK) model, which states that in an infinite system, the transformed fraction can be represented by:

$$f(t) = 1 - \exp\left[-\int_0^t J(\tau)V(t,\tau)d\tau\right] \tag{20}$$

where $J$ is the nucleation rate and $V$ is the extended volume of the transformed region. The extended volume here refers to the reversed domain volume as if there are no impingement and coalescence with other reversed domain volumes. Note that these general expressions do not impose any assumption on the nucleation and growth rates.

Later, to simplify the expression and transfer the model to ferroelectric systems, Ishibashi and Takagi adapted the JMAK model by adding two assumptions: 1) nucleation is either instant upon voltage applied, $J(\tau) = \delta(0)$, or at a constant rate, $J(\tau) = J_0$ and 2) the domain growth is at a constant velocity $v$, and therefore $V(t,\tau) = \gamma[v(t-\tau)]^d$, where $d$ is the dimensionality of domain growth and $\gamma$ is a geometric factor based on the geometry of the nuclei. Then, Equation 9 can be simplified into the KAI form:

$$f(t) = \frac{\Delta P(t)}{2P_s} = 1 - \exp\left[-\left(\frac{t}{t_0}\right)^n\right] \tag{21}$$

$t_0$ is the characteristic time of the polarization evolution and associated with the domain growth velocity $v$ and nucleation density $\rho$ ($\left(\frac{1}{t_0}\right)^n \sim \rho v^n$). Physically, it represents the average time for domain impingement. $n$ is the Avrami exponent that falls in the range of $d \leq n \leq d+1$, reflecting the mixing of the two nucleation categories, namely $n = d$ for instant nucleation and $n = d+1$ for constant nucleation rate. Thus, $n$ is viewed as an indicator of whether nucleation occurs concurrently during substantial domain growth. A cultural norm has been that a physical $n$ value should not be smaller than 1 or exceed 4, in accordance with the physical dimensionality of domain growth (1D to 3D). Switching transients with varying $n$ and $t_0$ are shown in **Figure 4a**. While in a linear time scale, all the profiles are in a similar sigmoidal shape with slightly different increase rate, the profiles in the logarithmic time scale indicate that only $n$ changes the slope and $t_0$ provides a shift in the time scale.

The KAI model is the most widely used model for its simplicity of only involving two fit parameters $t_0$ and $n$. Here we also use this model to illustrate how the dynamic error in conventional PUND method discussed in section 2.3.1 can affect the fit parameters. Still using the 10-$\mu m$ HZO capacitor as an example, fitting the original and corrected polarization transient to the KAI model results in $t_0 = 2.21\ ns, n = 1.86$ and $t_0 = 2.01\ ns, n = 1.72$ respectively, corresponding to 9% and 7.5% change in each (**Figure 4b**). Phenomenologically, the change in fit parameters relays a ~200 ps shorter domain impingement time and less abrupt switching. This discrepancy only accounts for the irreducible circuit effects and may become larger if parasitic elements are also considered. The correction of the conventional PUND analysis and the de-embedding of the measurement circuit need to be reinforced to reliably extract material parameters.

Furthermore, it should be reiterated that the KAI model strictly assumes that the nucleation and growth rates are to occur under constant voltage or in a regime that nucleation and growth is nearly independent of voltage such as extreme temperatures or extreme voltages. Any conditions that fail to match the assumption will lead to misinterpretation and likely unphysical results. We will discuss some cases later in this paper where time-varying voltage profile across the ferroelectric capacitor can result in diverge fit parameters, challenging the validity of most KAI-model-based fitting approaches.

**3.2 Classical nucleation and growth model: The NLS model**

Despite the KAI model providing a good fit to most single crystalline ferroelectrics, Tagantsev et. al. discovered that the polarization transients of the polycrystalline PZT film spans decades of time, which when plotting in the logarithmic time scale, cannot be described by varying $n$ in the KAI model. As in **Figure 4a** the range of $1 \leq n \leq 4$ only covers 1-2 decades. The "sluggish" transients were attributed to a distribution nucleation times in the material, $g(\lg \tau)$, reflecting that the nucleation events occur independently in individual regions or grains of the film. Because the nucleation time is relatively long, the time to fill a nucleated region (say a grain) becomes negligible. Hence the nucleation-limited switching. The switching transient described by this nucleation-limited mechanism is therefore given by:

$$f(t) = 1 - \int_{-\infty}^{\infty} \exp\left[-\left(\frac{t}{\tau}\right)^d\right] g(\lg \tau) \, d(\lg \tau) \tag{22}$$

Tagantsev et. al. assigned a flat distribution between wait time $\tau_{min}$ and $\tau_{max}$, with Lorentzian decay elsewhere. Jo et. al. [32]further developed the model, but instead of explicitly relating the distribution function to nucleation time distribution, a full Lorentzian distribution to the local field generated by dipole defects which pins the domain wall was assigned. Therefore, the distribution of nucleation times became

$$g(\lg \tau) = \frac{A}{\pi}\left[\frac{w}{(\lg \tau - \lg \tau_0)^2 + w^2}\right] \tag{23}$$

where $w$, $\lg \tau_0$, and $A$ are width, peak position, and normalization factor of the Lorentzian distribution. Strictly speaking, since the Lorentzian distribution of **Equation 23** was rationalized by a domain wall depinning mechanism, calling it a nucleation-limited mechanism is unfitting. However, due to the similarity of the expressions, following work adapts this modified expression and refers to it as a NLS model. Due to the new degree of freedom introduced by this distribution function, the growth dimension $d$ is usually recovered to be physical dimensions (integer 1, 2, 3) in the fitting of NLS model. By varying the width of the distribution linearly (from 0.2-0.8), the switching time spans 8 decades (**Figure 4c**). In the NLS model, the width is an empirical value without upper bound, but is believed to reflect pinning site concentration and applied voltage[32].

The NLS model captures the switching behavior of a variety of polycrystalline materials where defect concentration is high and domain wall motion is relatively localized. More recently, the NLS model has become especially popular after HfO$_2$-based ferroelectrics were discovered, enabling ferroelectrics to be readily integrable with CMOS processes[33].

Like the KAI model, there is no explicit voltage dependence in the NLS model but the freedom to choose nucleation time distributions or even fit arbitrary experimental polarization switching transients to extract the distribution allows for this nuance to be neglected.

### 3.3 Voltage dependent switching: Merz law

In 1954 Merz reported the observation of the relation between the switching time $t_s$ and applied electric field $E$ satisfies the equation

$$t_s = t_\infty \exp\left(\frac{E_a}{E}\right) \tag{24}$$

where $E_a$ is the activation field and $t_\infty$ is the terminal switching time[18]. Later, Miller and Savage[34] reported a similar relation in the domain wall propagation velocity $v$:

$$v = v_\infty \exp\left(-\frac{E_a}{E}\right) \tag{25}$$

Both equations are qualitatively equivalent when velocity is relatively constant throughout the process so that switching time scales with the inverse of velocity. Because exponential relations are very common in physical processes, especially when describing the probability of hopping over an energy barrier, these relations are widely accepted and interchangeably used to extract intrinsic activation field. The activation field ($E_a$) is a mean field that ferroelectric domain walls encounter when they propagate across an ensemble of locally disordered defects. Thus the domain wall sees a distribution of activation fields locally and the propagation shows fractal characteristics[35,36]. The pinning and depinning behaviors of ferroelectric domain walls simulate a disorder-controlled creep process due to their elastic nature. Hence, the domain wall motion described by **Equation 25** can be further explained by a creep formula[37–39]:

$$v \sim \exp\left[-\frac{U}{k_B T}\left(\frac{E_{C0}}{E}\right)^\mu\right] \tag{26}$$

where $U$ is a characteristic thermodynamic energy barrier, $E_{C0}$ is a critical depinning field, and $\mu$ is the dynamical exponent related to the nature of the defects. In most ferroelectric systems, $\mu \approx 1$ which relates **Equations 25** and **26** with $E_a = \frac{U E_{C0}}{k_B T}$.

Despite the disparate origins of the KAI and NLS models, the Merz law is found universal in both models. In the KAI model, the switching time $t_s$ (derived by $f \to 1$) is proportional to the characteristic time $t_0$. According to **Equation 24**, $t_0$ should then satisfy the Merz law. In the NLS model, the characteristic time $\tau_0$ is experimentally found to agree with the Merz law[16,32,40]. However, the post-hoc fitting of the characteristic time to the Merz law is under the critical assumption that the electric field is kept at a constant value $E$ throughout the switching process. As we will discuss later, the criterion is not always met, and the violation can lead to unphysical extrapolation.

### 3.4 Recent nucleation and growth model: The SNNG model

Wurtzite-structured ferroelectrics, namely the alloyed III-nitrides, have received extensive investigation as a newly discovered class of ferroelectric[41–44]. These new ferroelectrics are featured by a very square hysteresis loop with large coercive field and remanent polarization. The square hysteresis indicates abrupt polarization switching. Being a new class of ferroelectrics with these atypical quasi-static properties motivates the question of whether the switching dynamics of the wurtzite ferroelectrics are different from traditional ferroelectric materials. Although the answer is still under debate, various models seem to be applicable in independent studies[19,20,45,46] and less applicable in others. In some studies of the ferroelectric switching transients report long incubation times followed by a rapid switch once switching is onset. An application of the KAI model in this case would lead to very long $t_0$ (suggesting a long nucleation time) and an Avrami exponent exceeding 4[47]. To describe the trends, a new model, the SNNG model[17], was proposed and considers simultaneous nucleation and growth. In this model, the total number of nuclei is described by a sigmoidal curve

$$N(\tau) = N_0[1 - \exp(-\alpha\tau^m)] \tag{27}$$

Then a time-dependent nucleation rate is introduced into the JMAK model,

$$J(\tau) = \frac{dN(\tau)}{d\tau} = \alpha N_0 m \tau^{m-1} \exp(-\alpha\tau^m) \tag{28}$$

and the transient is then given by,

$$f(t) = 1 - \exp\left[-2\pi dv^2 N_0 \alpha m \int_0^t \tau^{m-1}(t-\tau)^2 \exp(-\alpha\tau^m)\, d\tau\right] \tag{29}$$

Here the exponent $m$ stands for the $m$-th order nucleation rate, whose value doesn't subject to any physical restriction and therefore can account for a large effective Avrami exponent. When $m > 1$, the nucleation rate here exhibits a peak shape, with constant $\alpha$ determining the nucleation peak position and $N_0$ the amplitude. When $0 < m \leq 1$, the nucleation rate is monotonically decaying (**Figure 4d**). While the peak shape and the decay can be explained by an increase in nucleation rate due to long incubation time followed by a depletion in the nucleation site, the physical meanings of the parameters $m$ and $\alpha$ are unclear. Moreover, the SNNG model also does not include an explicit voltage dependence, and how to implement one in this model is currently unclear. From the experimental data, the peak of nucleation rate seems to shift with applied field, indicating the voltage dependence should be imposed on $\alpha$, but analytical relation between $\alpha$ and the applied field was not investigated. However, the main idea of this model is that the concurrence of increasing nucleation rate and significant domain growth can potentially explain the seemingly unphysical Avrami exponent. The time-dependent nucleation rate is for the first time directly output from an analytical nucleation and growth model and qualitatively related to experimental observation by piezoforce response microscopy[48], implying additional variables can be used for complementary validation of modeling besides phase transformed fractions. The analytical form of decaying nucleation rates attributed to nuclei depletion can also be valuable to model kinetics in finite systems, as nucleation and growth models derived from statistical theories usually assume infinite systems.

### 3.5 Model misuse and Avrami exponent mystery

The above models either assume constant voltage during the switching process or provide no explicit voltage description. In contrary, as we discussed in session 2, the voltage waveform has skewed rising edge and is severely distorted by the polarization switching itself. The effect is dependent on both device sizes and supply voltages. Therefore, using the constant supply voltage in the model fitting, especially in microns-sized devices, can result in meaningless voltage dependence of the material parameters, because truly voltage-dependent parameters should vary with the instantaneous voltage instead of being held constant when applying the models.

Among the observations made from naive fitting of polarization transients, the Avrami exponent derived from the classical KAI model ($n_{KAI}$) evokes the most debate, as it has been extensively used as an indicator of nucleation mechanism (instantaneous or at constant rate) and growth dimensionality. The anomaly of static $n_{KAI}$ has motivated the emergence of the NLS (for $n_{KAI}$<1) and the SNNG (for $n_{KAI}$>4) models. These two models succeeded in recovering the physical dimension of the device (2D for thin films) and covering the validity range of KAI. In the NLS model, the transition of the kinetics is mainly accomplished by adopting the nucleation time distribution width. At high fields, the width becomes narrow and converges to a delta function which corresponds to the KAI kinetics[19,48]. In the SNNG model, a similar transition can be achieved by modifying the order *m* of the nucleation rate. However, these accommodations of the parameters have weak justification from the fundamental material properties and did not consider the effect of time-dependent voltage. Another angle of the Avrami exponent is the dynamic behavior. If a generalized definition of the Avrami exponent is taken, i.e. $n = \frac{d\ln(-\ln(1-f))}{d\ln t}$, the experimental data will give a dynamic *n* value[36,49]. In fact, the NLS and SNNG models capture this dynamic behavior (**Figure 5a**) due to the introduction of time-dependent nucleation rates, an aspect that is overlooked in the previous investigations. They reveal that the relative time scale of nucleation and growth is an intrinsic reason for the dynamic Avrami exponent.

Here we introduce a different perspective to resolve the Avrami exponent mystery. The Avrami exponent is an effective growth dimensionality that reflects the expansion pattern of the domains, and thus is not necessarily restricted by or related to the physical dimension of the device. If the expansion of the domain is significantly hindered by the microstructure, internal field, and the finite size effects in the device[50–52], the domain wall cannot propagate isotropically in all directions within the physical dimension. Deceleration and deformation of the domain shape happen and lead to a lower effective dimension, which can also account for the case of Avrami exponent below 1 (**Figure 5b**). This case often appears as unobservable domain growth while nucleation of new domains is prominent.

On the other hand, high growth dimensionality is convoluted with a circuit effect, where the ramp of voltage potentially synchronizes the spike of nucleation and growth rate. For example, in **Figure 5c,d**, the switching transients under voltage pulses with different nominal rise times (~50 ps and ~1 ns) and supply voltages (1 V and 2 V) are measured on a BaTiO$_3$ (BTO) capacitor. By deliberately applying these voltage profiles, we obtain distinct Avrami exponents of $n_{KAI}$ of 2.1 and 5.4, the latter considered unphysical in the conventional perspective. By observing the polarization transients, lower $n_{KAI}$ is associated with a more sluggish rise in polarization and a shorter wait time to reach measurable polarization. Accordingly, we posit that in the case of high $n_{KAI}$, the slow ramp of voltage keeps the nucleation and growth indiscernible initially. The dynamic $n$ values also exhibit different profiles, which are not implicated by the static $n_{KAI}$, but can be explained the ramping of voltage and the consequent acceleration in nucleation and growth rate. Once the voltage reaches a certain threshold, nucleation and growth accelerate to a considerable speed concurrently and leads to a steep increase in the switched domain volume. Our experimental data indicates that when the rise time of the voltage pulse and switching time of the device is comparable, unphysically high Avrami exponent can occur even in a traditional ferroelectric material. This possibility remains in the case of the reported abrupt switching in doped-AlN ferroelectrics[47], where larger devices of 100s-ns switching time were driven by ~200 ns rise time pulse generator. Therefore, analytical models need to incorporate time-dependent nucleation and growth rate based on the instantaneous voltage profile to properly interpret the dynamic growth dimensionality. Additionally, the dynamic Avrami exponent can be an effective way to validate the model.

**3.6 Outlook for nucleation and growth modeling**

On top of the fundamental physics, voltage dependent nucleation and growth models are important for circuit modeling. Ferroelectric materials are promising candidates for in-memory computing and neuromorphic devices, where the programming and readout of the polarization states rely on highly versatile waveforms, such as sinusoidal oscillations, pulse trains of varying amplitudes, widths and numbers. Analytical models that accurately predict ferroelectric response need to be integrated into circuit-level modeling, to optimize operation waveforms, compute energy consumptions, and estimate readout signal as well as variations etc. on large-scale arrays. Currently circuit modeling are mainly based on phenomenological models such as Preisach model[53], Landau model[54–56], and Jiles-Atherton model[57]. These models borrowed the theories from ferromagnetic switching for their capability of capturing hysteric behavior based on the full history of the applied field. However, these models cannot produce realistic time domain transients. The time scale in these models is arbitrary as their parameters are purely empirical without physical connection to nucleation and growth kinetics. As a consequence, the time step in circuit modeling has to match the one in the PUND measurement set up.

On the other hand, the current frameworks of KAI, NLS and SNNG models built on the assumption of constant voltage during switching do not allow the insertion of a path dependent voltage profile, preventing their application in circuit modeling[49]. For KAI and NLS, usually the characteristic time is considered to have voltage dependence according to Merz law. Equivalently, the formula for transformed fraction $f$ will become $f(t) = F\left(\exp\frac{E}{E_a}\right) = F(E)$. This substitution eliminates the path dependence of the voltage, implying that the ferroelectric material has no memory of the prior waveform and that it reaches the same state as the instantaneous voltage reaches the same value. This behavior is against the hysteric nature of ferroelectric memory and can result in unphysical switching curves. For SNNG model, if voltage dependence is enforced on the parameter $m$ and $\alpha$, which reside in the integrand, technically the function captures voltage path dependence. Yet the physical meaning and hence their relation to electric field is ambiguous.

Moving forward, tethering instantaneous voltage profile to nucleation and growth kinetics is crucial for analytical models to function beyond fitting. A proposed scheme to leverage a nucleation and growth model is shown in **Figure 6a**. Polarization transients under real voltage waveforms are simultaneously input into the model to extract physical parameters. These parameters are intrinsic to the material regardless of external fields and thereby allow the prediction of ferroelectric response driven by arbitrary waveforms. Typically, the scalability of technologies requires sub-micron-sized devices which are in favor of preserving the defined shape of the waveform. Still in non-ideal measurement schemes, the model can output the time-dependent capacitance of the ferroelectric and allow for computing the waveform distortion based on the circuit impedance and current transients.

Recently, a dynamic-field driven nucleation and growth (DFNG) model based on this principle has been developed with voltage dependent nucleation rate and domain wall velocity considered[49]. Using this model, a single set of intrinsic material parameters is able to fit ferroelectric switching transients under different supply voltages and time-varying waveforms (**Figure 6b**). While capturing the smooth sigmoidal shape of the polarization evolution, the current profiles replicate the small oscillation in the voltage profiles as well. With such a sensitivity to the time dependent variation in voltage, it is promising for generating realistic circuit response. For example, potentiation curves for ferroelectric neuromorphic devices can be simulated (**Figure 6c**) amongst other dynamic measurements (like $C(V(\omega))$ and $P(E(\omega))$ in ref 49). By changing the programming pulse amplitudes, the linearity of multi-level states can be optimized. This simulation demonstrates a potential practice of nucleation and growth model assisting the development of ferroelectric technologies.

## 4. Conclusion

Investigation of ferroelectric switching kinetics via electrical measurements are increasingly moving toward the intrinsic limits of materials in terms of switching speed and device scaling. The central objective is to elucidate and ultimately control the underlying mechanisms of polarization reversal, governed by the nucleation and growth of domains. However, ferroelectric capacitors are inherently operated within a measurement circuit, making their interplay with external circuit elements unavoidable. Whether experimentally extracted materials parameters are free from circuit-induced artifacts, and whether conventional analytical models remain valid in the ultrafast switching regime hasn't been critically considered. In this paper, we aim to highlight and quantify some of the overlooked issues. We show that ferroelectric capacitors are always operated under highly distorted voltage waveforms due to the significant displacement current generated during polarization switching. Such distortion is dependent on capacitor size and supply voltage. This necessitates the monitoring of real-time voltage directly across the ferroelectric capacitor and a modification to the classical PUND analysis, in order to accurately extract the intrinsic ferroelectric response. Specifically, individual values of the circuit elements need to be measured, including the linear component of the ferroelectric capacitor and parasitic elements introduced by device design and fabrication. Furthermore, the time-dependent voltage profile challenges the applicability of traditional nucleation and growth models for the extraction of material parameters. Under rapidly varying voltage, intrinsic material parameters independent of voltage should be used in a model, in contrast to the conventional practice where characteristic time scale as a parameter is dependent on the supply voltage via Merz law. The voltage-dependent proxies should be nucleation rate and domain propagation velocity instead, which can be cross-validated by complementary experimental techniques such as piezoforce microscopy and electron microscopy, as well as by theoretical calculations. Overall, this paper seeks to provide a framework for rigorously extracting intrinsic switching transients and for developing and best utilizing analytical switching models to facilitate ferroelectric studies and technological advancements.

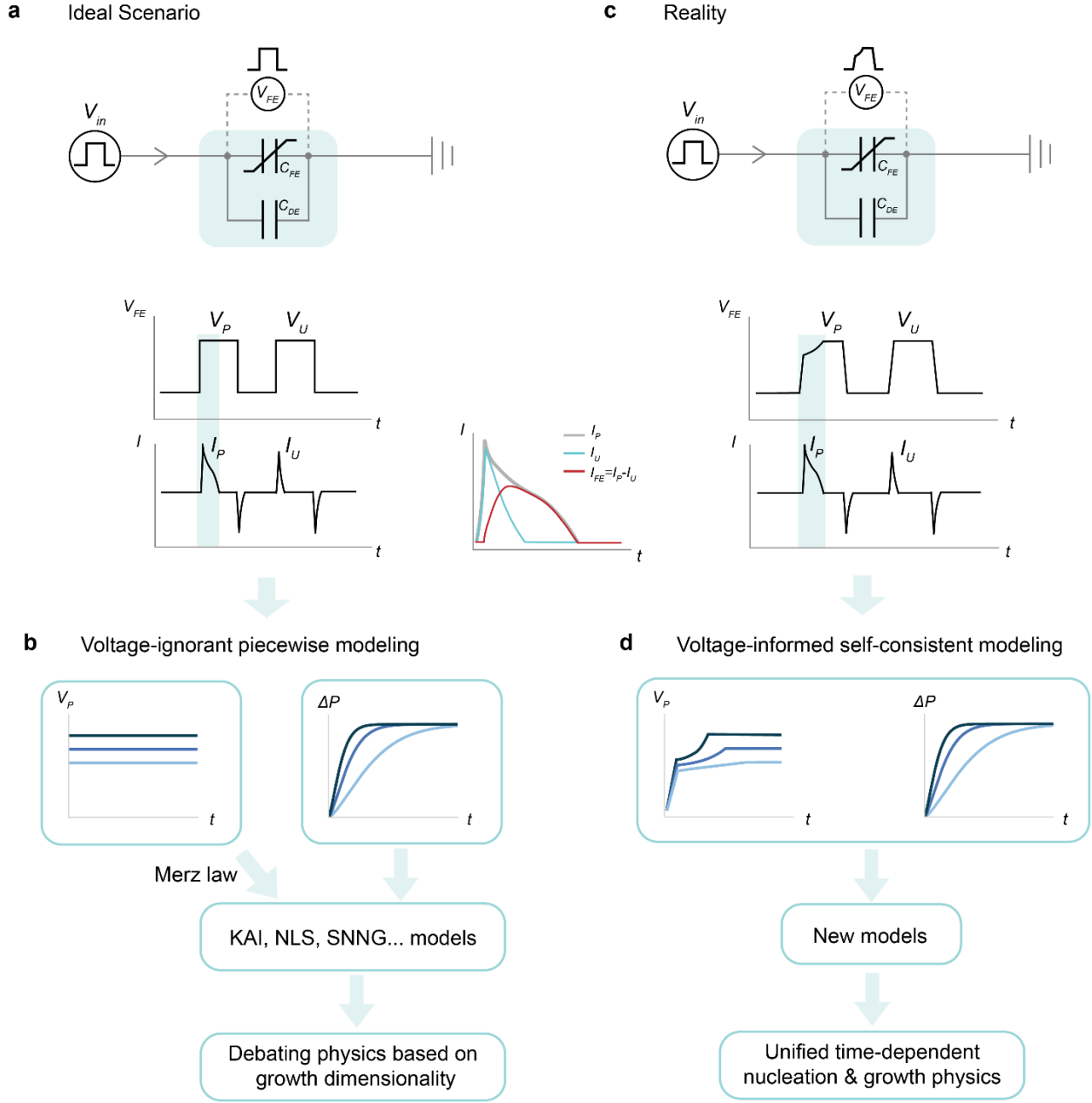

**Figure 1. Idealities and realities of the PUND methodology in ferroelectric switching transient studies. a,** An ideal measurement and analysis scheme commonly assumed in the literature. The voltage ramps to the set value instantaneously and the ferroelectric capacitor causes no distortion to the square pulse. The positive-up-negative-down (PUND) method is used to extract the ferroelectric switching current ($I_{FE} = I_P - I_U$) and polarization transients ($\Delta P(t)$). **b,** The polarization transients are then fit to a prevalent nucleation and growth model such as KAI or NLS

model. Parameters from these models usually obey Merz law under the constant voltage assumption. However, using these models to interpret data where the voltage varies in the transient, either directly or by direct substitution of the Merz expression into the model, leads to misquantification and misinterpretation of physical parameters. **c**, A real measurement circuit encounters problems of waveform distortion due to the finite rise time of the signal generator, the nonlinear response of the ferroelectric capacitor, and potential interplay with other circuit elements. This makes the profiles of $V_P$ and $V_U$ inequivalent and breaks the key assumption of the PUND method. **d**, Because of the time-dependent voltage profile, the PUND method needs revision or a new nucleation and growth models is needed to handle the fundamental analysis of the polarization transient. Such methods would then extract intrinsic material physics irrespective of the voltage across the ferroelectric capacitor.

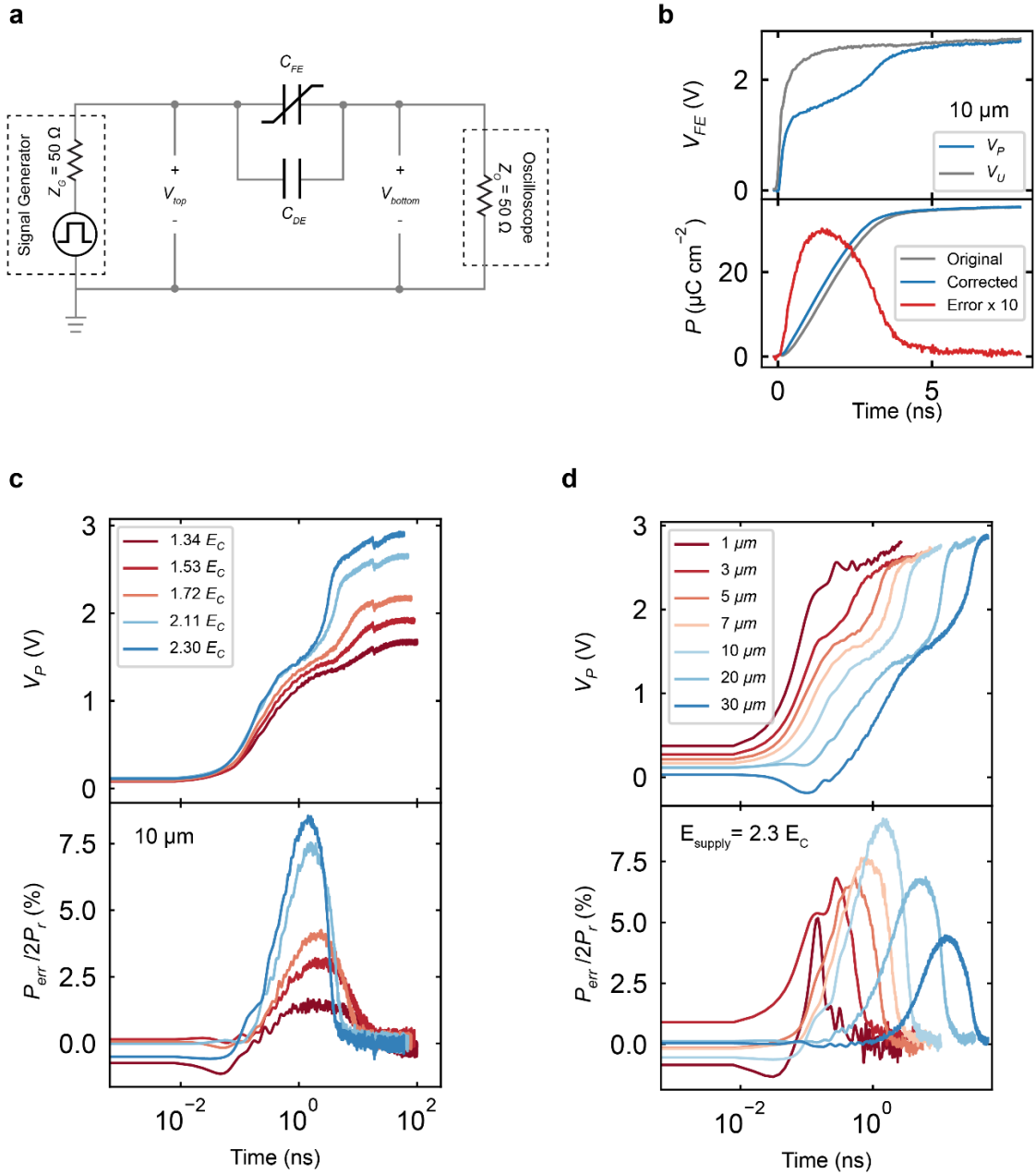

**Figure 2. Quantification of errors in ideally assumed PUND measurements. a,** Schematic of the measurement circuit with the actual voltage across the ferroelectric capacitor monitored. The voltage during P ($V_P$) and U ($V_U$) pulses can then be calculated by $V_{top} - V_{bottom}$. **b,** Analysis performed on a 10-μm HZO capacitor. Notably, the P pulse and U pulse are not identical, leading to error in transient polarization calculated by conventional PUND methods. **c,d,** Supply voltage-dependent (**c**) and capacitor-size-dependent (**d**) distortion of P pulse waveforms and errors in transient polarization.

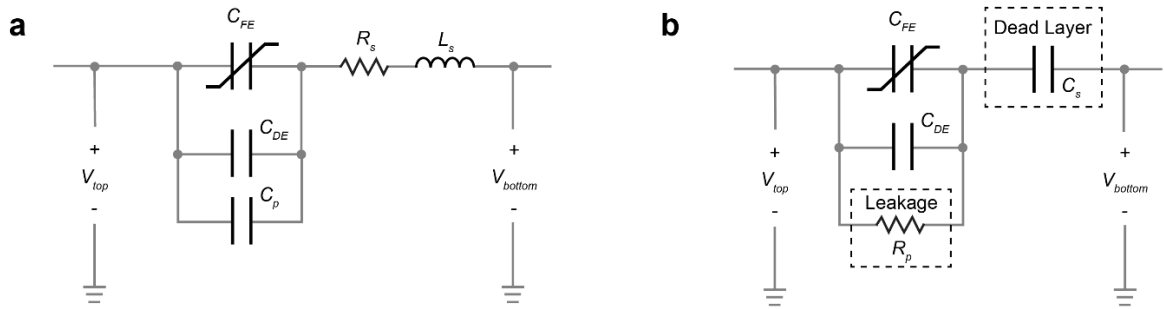

**Figure 3. Possible linear components embedded in the MFM structure. a,b,** Equivalent circuit diagrams of the measured MFM structure when parasitic circuit components (parallel capacitor, series resistor and series inductor) (**a**) and material deficiencies (dead layer and leakage) (**b**) are present.

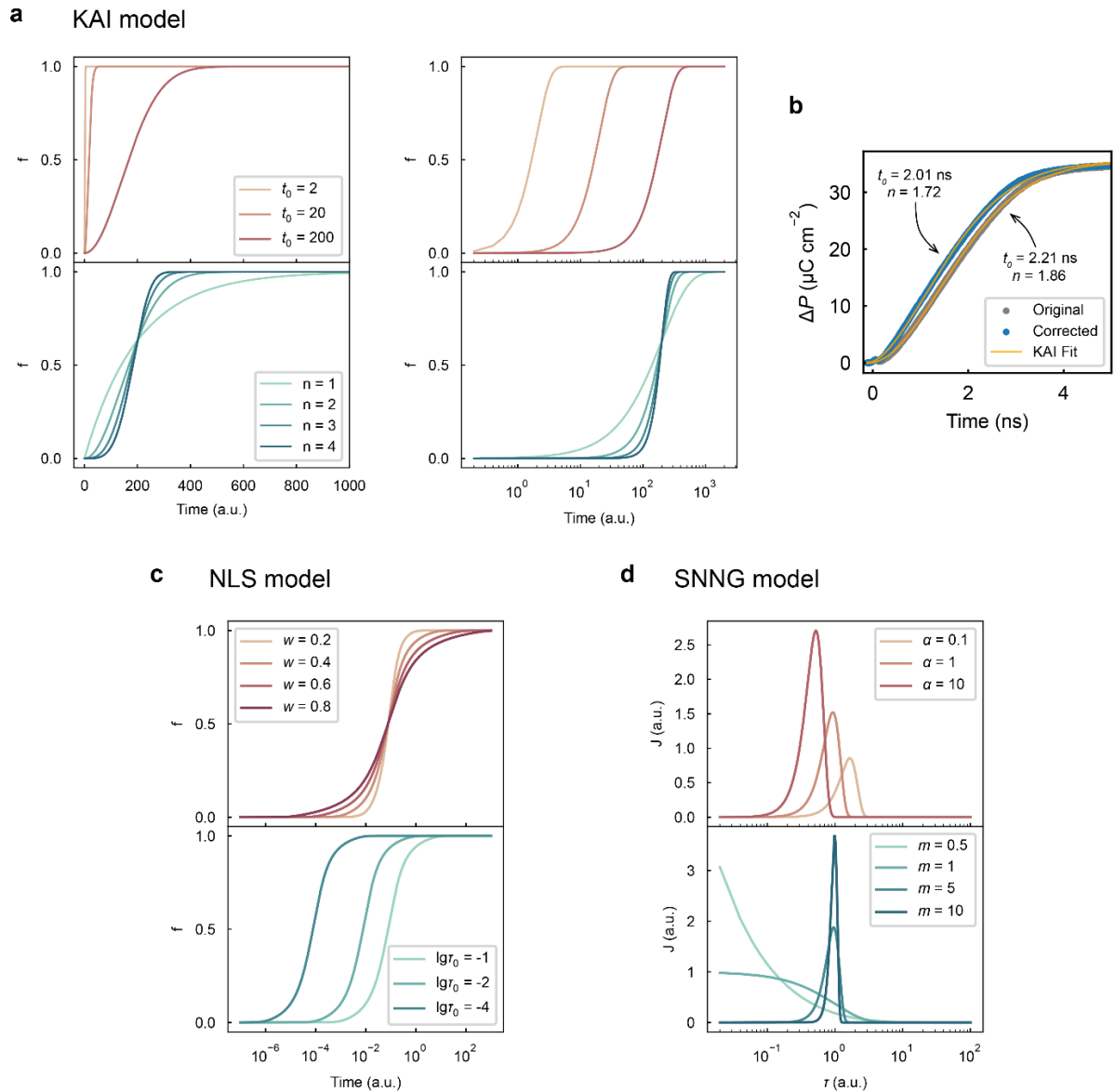

**Figure 4. Behavior of constant voltage nucleation and growth models. a,** Polarization transients simulated with KAI model (linear (left) and logarithm (right) time axes) by varying model parameters. **b**, Comparison of KAI model fit (orange) to polarization transients obtained by conventional (grey) and corrected PUND methods. **c,d,** Polarization transients simulated with NLS (**c**) and nucleation rate simulated by SNNG (**d**) models under variance of key model parameters.

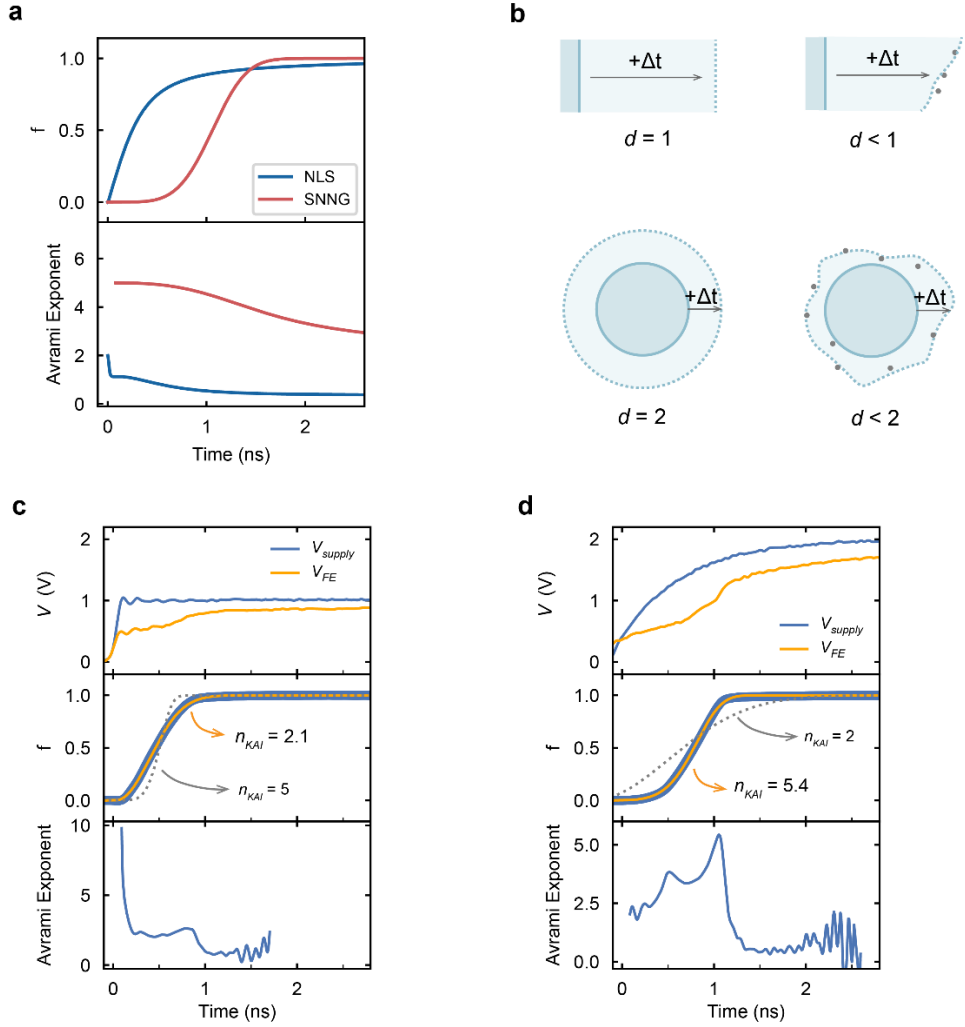

**Figure 5. Avrami exponent mystery. a,** The NLS and SNNG models explain unphysical Avrami exponents via different mechanisms. By using a time-dependent nucleation rate, they capture the dynamic behavior of the exponent. **b**, A schematic of the alternative explanation for growth dimension lower than physical dimension. It happens when the elastic domain wall is deformed by disordered defects, and therefore the propagation does not follow the trajectory dictated by the physical dimension. **c,d,** Different pulses, either by rise time or amplitude, can produce distinct Avrami exponents even on the same ferroelectric capacitor. The Avrami exponent obtained by KAI model can exceed 4; When the polarization evolves with time-varying voltage, the dynamic Avrami exponent ($\frac{d\ln(-\ln(1-f))}{d\ln t}$) can become large and unphysical.

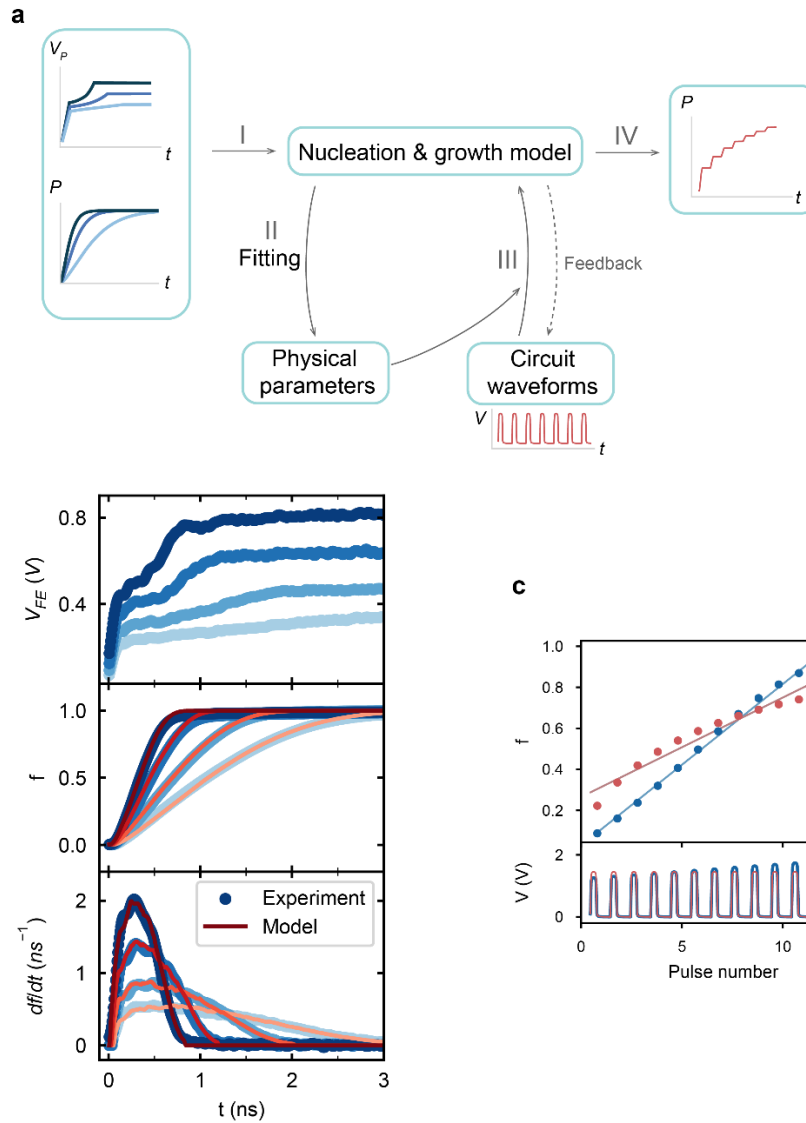

**Figure 6. Perspectives on nucleation and growth models. a,** Compact analytical nucleation and growth model should fulfill the task of fitting to polarization transients under time-varying voltage waveforms with intrinsic material parameters. The extracted parameters in turn should enable simulation of realistic circuit responses based on input arbitrary waveforms. **b**, DFNG model fit to experimental switching transients under time-dependent voltage waveform, fulfilling step I and II in (**a**). Reproduced from ref. 49. **c,** An example utilizing the recently developed DFNG model to generate a potentiation profile in a ferroelectric-based neuromorphic computing scheme, fulfilling step III and IV in (**a**). The simulation predicts that by varying the amplitude of each potentiation pulse, the linearity in multi-level operation is improved (blue) compared to constant amplitude (red).

## Acknowledgements

This work is supported by the ONR grant N000142612047.

## Data Availability Statement

All data that supports the findings of this study are available from the corresponding author on reasonable request.

A meme summarizing part of this work can be found on https://ferroelectronicslab.com/memes/